# IHCV: Discovery of Hidden Time-Dependent Control Variables in Non-Linear Dynamical Systems

Juan Munoz[1], Subash Balsamy[1], Juan P. Bernal-Tamayo[1], Ali Balubaid[1], Alberto Maillo Ruiz de Infante[1], Vincenzo Lagani[1,4,5], David Gomez-Cabrero[1,3], Narsis A. Kiani[6,7], and Jesper Tegner[1,2,7,8,*]

[1] Biological and Environmental Science and Engineering Division, King Abdullah University of Science and Technology (KAUST), Thuwal 23955-6900, Saudi Arabia.

[2] Computer, Electrical and Mathematical Sciences and Engineering Division, King Abdullah University of Science and Technology (KAUST), Thuwal 23955-6900, Saudi Arabia.

[3] Translational Bioinformatics Unit, Navarrabiomed, Complejo Hospitalario de Navarra (CHN), Universidad Pública de Navarra (UPNA), IdiSNA, Pamplona, Spain.

[4] SDAIA-KAUST Center of Excellence in Data Science and Artificial Intelligence, Thuwal 23952, Saudi Arabia

[5] Institute of Chemical Biology, Ilia State University, Tbilisi 0162, Georgia

[6] Algorithmic Dynamic Lab, Department of Oncology and pathology, Karolinska Institute, Stockholm, Sweden.

[7] Unit of Computational Medicine, Department of Medicine, Center for Molecular Medicine, Karolinska Institutet, Karolinska University Hospital, L8:05, SE-171 76, Stockholm, Sweden

[8] Science for Life Laboratory, Tomtebodavagen 23A, SE-17165, Solna, Sweden

[*] Correspondance: jesper.tegner@kaust.edu.sa

# ABSTRACT

Discovering non-linear dynamical models from data is at the core of science. Recent progress hinges upon sparse regression of observables using extensive libraries of candidate functions. However, it remains challenging to model hidden non-observable control variables governing switching between different dynamical regimes.

Here we develop a data-efficient derivative-free method, IHCV, for the Identification of Hidden Control Variables. First, the performance and robustness of IHCV against noise are evaluated by benchmarking the IHCV method using well-known bifurcation models (saddle-node, transcritical, pitchfork, Hopf). Next, we demonstrate that IHCV discovers hidden driver variables in the Lorenz, van der Pol, Hodgkin-Huxley, and Fitzhugh-Nagumo models. Finally, IHCV generalizes to the case when only partial observational is given, as demonstrated using the toggle switch model, the genetic repressilator oscillator, and a Waddington landscape model.

Our proof-of-principle illustrates that utilizing normal forms could facilitate the data-efficient and scalable discovery of hidden variables controlling transitions between different dynamical regimes and non-linear models.

The scientific and machine learning community is increasingly joining forces to accelerate and enable scientific discovery using data-driven techniques [1]. Such a pursuit goes beyond data analysis and pattern detection. In science, we need to answer what-if questions about controlling and steering a system between different dynamical regimes [2,3].

Broadly, most efforts to date rely upon strong, potentially unrealistic assumptions to identify models from observations[4,5]. These efforts include linear modeling, assuming sparse linear models with few non-zero coefficients, despite that phenomenon in nature having a wide variety of dynamics, which in turn requires non-linear differential equation models[6]. Moreover, several methods assume the availability of a large amount of observational training, experimental or synthetic data, often requiring perturbation data[7,8]. However, we rarely find such rich observational data with or without perturbations when studying natural systems. Instead, a library of functions coupled with constraints such as sparseness or parsimony has been used to meet the challenge of decoding non-linear systems[5,9]. Yet, these approaches either scale poorly with increasing basis functions or require strong assumptions of only using a combination of a few terms to represent the data.

A non-linear dynamical system can readily exhibit complex switching dynamics alternating between fixpoints, oscillations, or chaotic dynamics [10–12]. However, discovering the core control mechanisms governing such rich flora of dynamical fingerprints remains challenging. Furthermore, we cannot assume that we can access all relevant variables to perform such identification. For example, data is, as a rule, incomplete due to the sparsity of measurements, and we may not even know the unknown variables which are part of the physical system of interest. This formulation has broad applications in earth science/climate modeling, neuroscience, genomics, stem cell biology, and developmental biology[10] (Fig. 1A). In these areas, we do not, as a rule, have access to fundamental models formulated from first principles. Thus, the problem of model discovery is intrinsically linked with the challenge and existence of such hidden variables [13,14]. Here we develop a method IHCV to Identify a Hidden Control Variable from temporal observations of a non-linear dynamical system without requiring perturbation data. Our key idea is to use normal forms from dynamical systems theory as minimal model building blocks [15]. As a result, we model the transitions between different dynamical regimes as the system dynamics evolve according to unobserved control variables.

Inspired by dynamical system theory [15], we address this problem of modeling complex dynamical systems and their transitions between different states using normal forms as fundamental model building blocks[16,17]. To assess the feasibility of such a concept, we use the pitchfork, the transcritical, the saddle node, and the Hopf bifurcations as proof-of-principle test models (Fig. 1B). To illustrate, consider that we observe and measure one and only one variable (x), which we discover since it at some point in time starts to grow over time (Fig. 1C). Therefore, there is another process that we do not observe which drives this increase in x or, equivalently, shifts the value of the fixpoint for x. We model our observation using the saddle-node normal form $\dot{x} = -\alpha + x^2$. Importantly, the parameter $\alpha$ is hidden and is a function of time ($\alpha(t)$). Our task is to reconstruct the temporal dynamics of the hidden variable $\alpha$ from the observed data (x). To solve this, we approximate the system to be near a stable manifold. This amounts to saying that the dynamics of the control variable are sufficiently slow such that the induced shifts in x allow the system to evolve in the neighborhood of its stable manifold. As a result, this reduces the problem of finding a polynomial representation of $\alpha$ over the observed temporal interval. Here we model $\alpha$ such that $\alpha(t) = at + bt^2 + ct^3$ and solve for the weights a, b, and c. This enables a reconstruction of alpha, sufficient to account for the observed dynamics of x (Fig. 1C). Next, we asked whether this proof-of-concept generalizes to a broader class of models. To this end, we evaluated and benchmarked our method using models such as normal form, classical dynamical systems, neural systems, and models in population biology, synthetic biology, stem cell biology, and developmental biology (Fig. 1A).

First, we evaluate IHCV using transitions between different dynamical regimes, such as between two different fixpoints in normal-form models. We generated observational temporal data such that a low-steady-state switched to a high-steady-state level after some time (Fig. 2a, left). The IHCV method solves for a parametric representation of the hidden driving variable (Fig. 2a, right) driving this jump using the pitchfork model (Fig. 2a, middle). Using the saddle-node model solved the case with a growing fixpoint (Fig. 2b). Next, we addressed the challenge when an oscillation is destabilized, such that a fixpoint emerges in the system (Fig. 2c, left). Using a Hopf normal form model (Fig. 2c, middle), our method could solve for a parametric representation of the hidden driving variable (Fig. 2c, right). Similarly, we could solve corresponding cases for transcritical and pitchfork bifurcations (Supp. 1).

The dynamical systems literature has extensively studied the Lorenz and van der Pol models [15,18]. Here we took a different path asking whether we could identify from observation data a hidden variable that, in these systems, alters the frequency and amplitude of their respective oscillatory behavior in a time-dependent manner. For the Lorenz attractor, we generated data with a smooth increase of the amplitude and frequency over time (Fig. 2d, left). As time progressed, we slowed down the relaxation oscillator for the van der Pol (Fig. 2e, left). In both cases, we could identify the time-dependent dynamics of the hidden variable (Fig. 2d,e, right) using the respective model (Fig. 2d,e middle) in conjunction with the observation data.

The Hodgkin-Huxley is an iconic model for node dynamics in modeling biological neural networks [19]. Here we used the external current injection as a hidden control variable, and we could reconstruct the hidden dynamics of the current injection (Fig. 2f, right) from the observational data Fig. 2f, left).

Next, we validated these observations (Supp. 2) using the classical reduced two-dimensional Fitzhugh-Nagumo [19] neural (FHN) model (Fig. 2f). Here, the IHCV method was challenged to discover the hidden drive destabilizing the fixpoint in the FHN model into a subsequent emergence of relaxation oscillations and the return to another growing fixpoint. Finally, we further validated the IHCV method using data from different dynamical regimes produced by the Lotka-Volterra prey-predator model [18] for population dynamics (Supp. 3).

In all these proof-of-concept demonstrations, we have recovered the temporal dynamics of a hidden variable. Next, we asked whether IHCV generalizes to the more challenging scenario where we cannot access the domain-specific model equations. This concern is relevant in synthetic, stem cell, and developmental biology (Fig. 1A). First, we used the genetic toggle switch model[20], which models two stable gene expression states in a coupled two-gene circuit. This engineered synthetic biology gene circuit captures the experiment where an experimental induction switches the system from one stable to another stable state. As input (Fig. 2h, left) to IHCV, we have an upper/higher steady state, which switches into the low state after some time. IHCV successfully reconstructs the qualitative dynamics of the hidden driving variable (Fig. 2h, right) without access to the model equations. Importantly, while we do not assume access to the "true" model equations, the fit does not have a zero mean-square error. Yet, IHCV captures the monotonic increase in the hidden variable. The synthetic repressilator circuit [21] consists of three interconnected genes that produce an oscillatory system. We drive the circuit from a fixpoint into an oscillatory mode with increasing frequencies. Using only these observational data (Fig. 2i, left), without access to the model equations, IHCV identifies the qualitative dynamics of the hidden control variable (Fig. 2i, right).

Cellular reprogramming is one of the critical discoveries in stem cell biology[22]. Stem cells are pluripotent and can be transformed into other distinct cell types. Cells occupy, in this sense, different positions in a developmental landscape, technically referred to as a Waddington landscape[23]. Such landscapes have also been used to account for the development of different cell types and organs, resulting in an organism. Several factors, in part unknown, contribute to the cellular progression along such a Waddington landscape[10,24]. Here we ask whether our method could uncover such a hidden driver by observing a system transitioning from one stable state into two new stable states (Fig. 2h, left). Using a supercritical pitchfork model (Fig. 2j, middle), IHCV recovered the temporal dynamics of the hidden driver variable (Fig. 2j, right).

Here we have developed a data-efficient derivative-free technique for model discovery, identifying the hidden control variable with unknown dynamics and steering the system between different dynamical regimes. We have used normal forms as universal, scalable model building blocks to capture the sample observational data near a slow manifold. The problem of modeling a dynamical system is transformed into selecting a bifurcation normal form with a similar trajectory on the stable manifold. Such a stable manifold can be transformed using a convenient dynamic for the bifurcation parameter. In other words, we can generate a rich dynamic based on a single system by only manipulating the dynamic bifurcation parameters. This technique works for many models ranging from normal forms as a proof-of-principle, including Lorenz and the van der Pol models, to dynamical systems models in population biology, neuroscience, genetics, and stem cell biology. Future work includes scaling up this proof-of-principle for larger systems to enable large-scale mechanistic modeling [25].

**FIGURE CAPTIONS**

**Figure 1:** The IHCV Method: **a)** The domain of applicability includes climate modeling, neuroscience, genomics, and stem cell biology. **b)** Hidden driver variable (blue line) switches the system dynamics between different dynamical regimes illustrated by the pitchfork system (upper red line) and a Hopf bifurcation system (lower red line). **c)** Illustration of the IHCV method using the saddle-node bifurcation. Upper panel: the hidden driver variable α (x-axis) controls the observed state variable (y-axis). Lower panels: collecting data of the observed state-variable over time (left) is sufficient to recover the dynamics of α (right) using polynomial regression solving for the weights a, b, and c such that $\alpha(t) = at + bt^2 + ct^3$.

**Figure 2.** Learning hidden control from observations. Qualitative dynamics (first column), and the second column displays the raw observational data points. The generative model (third column) predicts the hidden variable's dynamics (fourth column). The dotted blue line marks the learned dynamics, whereas the red line indicates the actual dynamics. The first three models **(a-c)** are based on the theory of normal forms, including pitchfork, saddle-node, and Hopf bifurcation. The Lorentz and Van Der Pol physics models **(d-e)**. The remaining two models from neuroscience **(f-g)** are the Fitzhugh-Nagumo and Hodgking-Huxley. Finally, in the last three models **(h-j)**, the qualitative dynamics of the model using a library of bifurcation normal forms are predicted when we do not have access to the fundamental equation. The normal forms adequately capture the system's dynamics, despite a discrepancy in column 4 for h-j between the actual and learned values.

## MATERIALS and METHODS

The methods are uploaded in a separate file. In addition, the datasets, models, and codes are shared on a dedicated GitHub page https://github.com/munozdjp/IHCV including worked-out data examples.

## DATA AVAILABILITY

All data and software are publicly accessible, as detailed in the Materials & Methods section.

## FUNDING

The King Abdullah University of Science and Technology supported this work.

## CONFLICT OF INTEREST

The authors declare no conflict of interest.

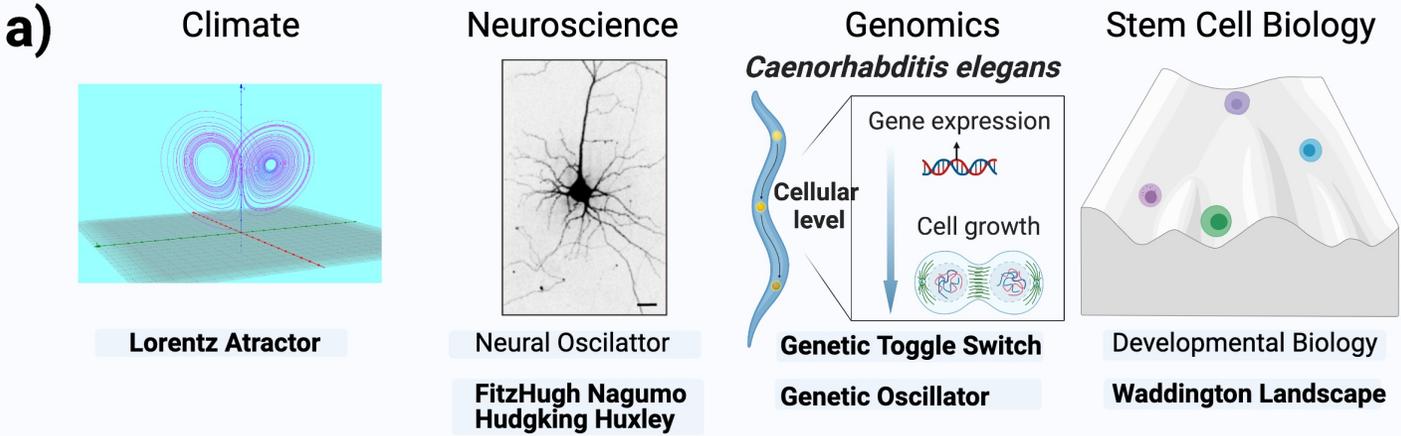
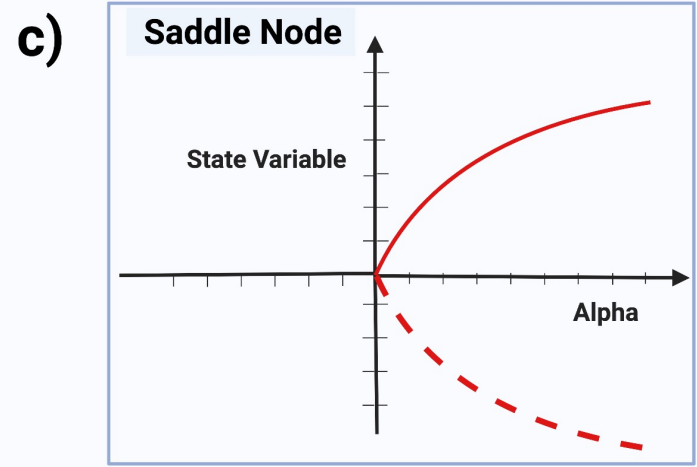
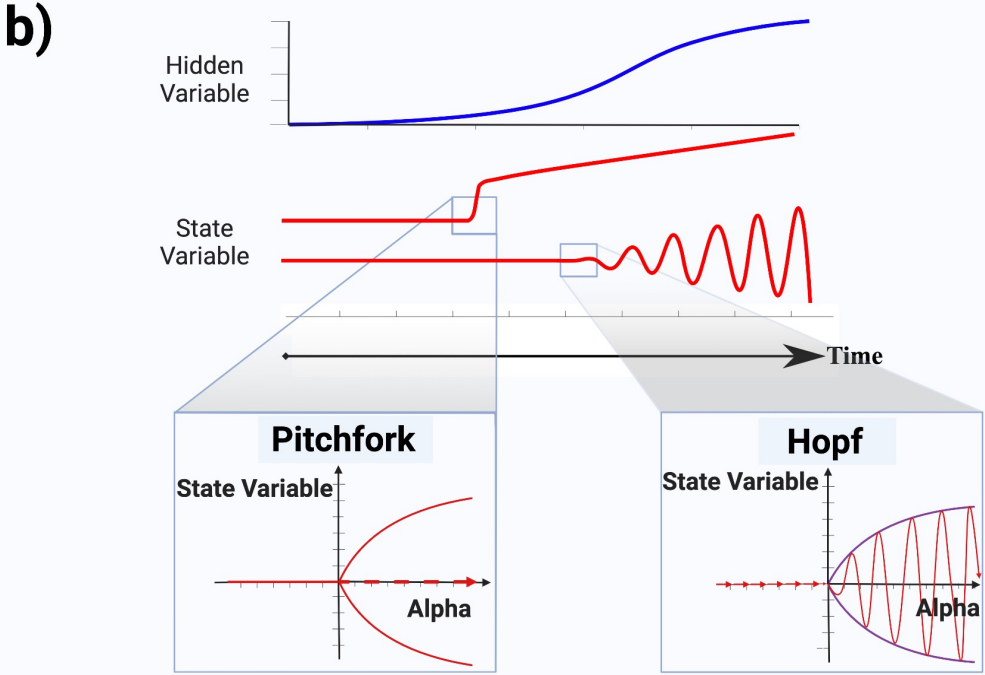
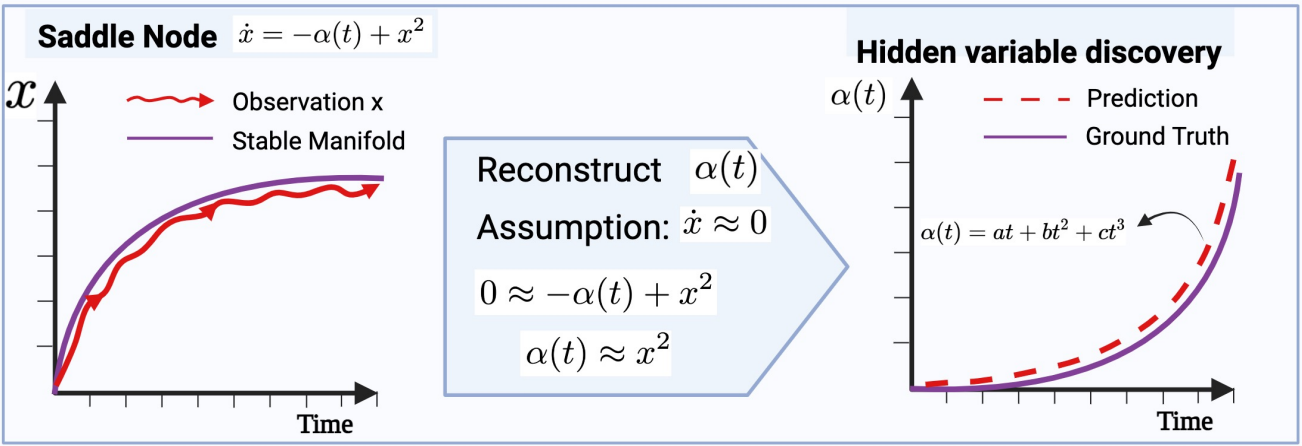

Fig. 1

Fig. 2

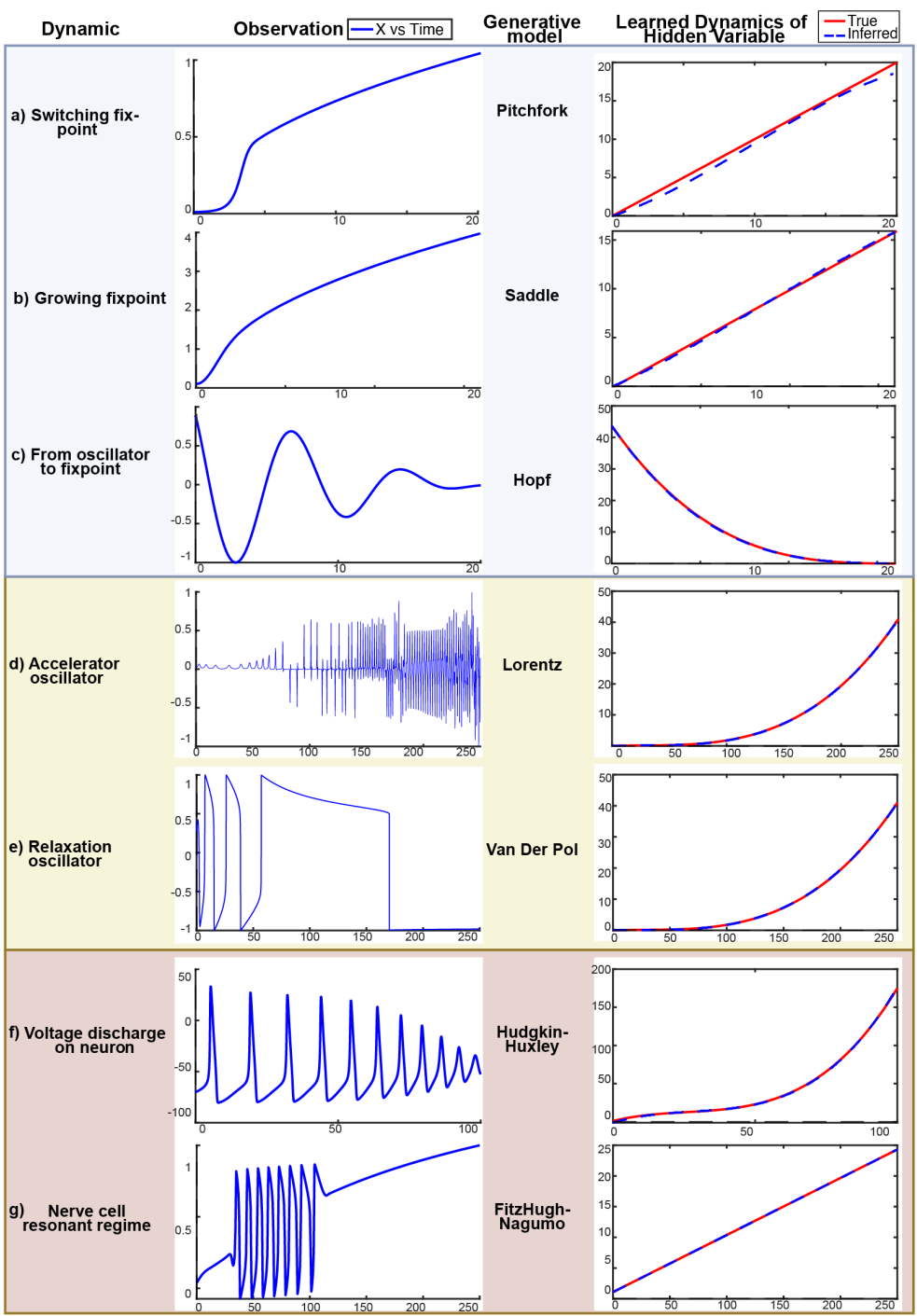
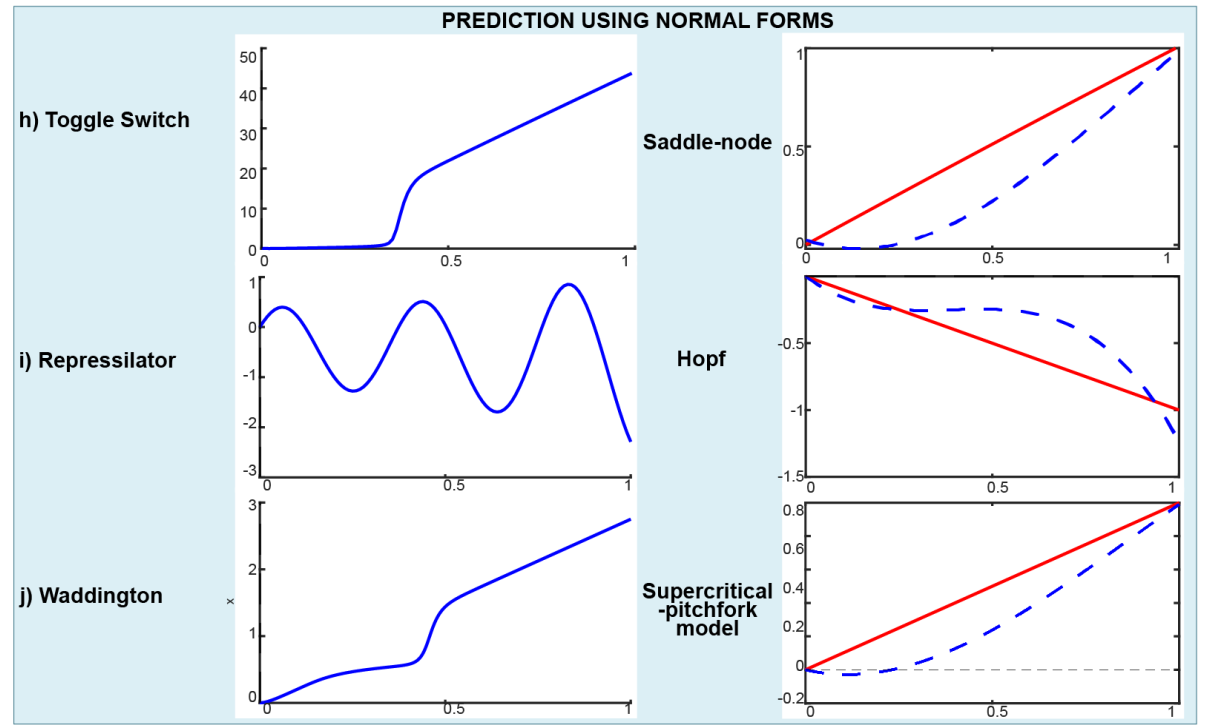